\documentstyle[12pt]{article}
\title{On the spin orbit force in the Wilson loop context}

\vspace{2cm}

\author{V.M. Kustov \thanks{e-mail:kustov@vxitep.itep.ru}
\\ Institute of
Theoretical and Experimental Physics\\ 117259, 
Moscow, Russia}
\date{}

\begin{document}
\maketitle

\begin{abstract}
The Green function of the quark-antiquark system in the confining 
background field is analysed using the Feynman-Schwinger formalism. The 
Hamiltonian for the case of massive spinning quarks is obtained in the 
form containing essentially nonhermitian part. The eigenvalue problem 
for such type of the Hamiltonian is discussed, and it is shown that no 
complex eigenvalues arise. The corresponding nonunitary Foldy-Wouthuysen 
transformation  is performed to obtain the hermitian Hamiltonian, and 
the standard spin-orbit interaction term is recovered. 
\end{abstract}

\vspace{1.5cm}

\newpage

\section{Introduction}
%____________________________________________________________________________
The most natural way to discuss quark confinement at the constituent level 
is in terms of Wilson loop

\begin{equation}
W(C)=Tr~P~exp~ig\oint A_{\mu}^a\lambda^a dz_{\mu}.
\end{equation}

The area law asymptotics for the Wilson loop (1) averaged over the vacuum 
gluonic field

\begin{equation}
<W(C)>\sim exp (-\sigma S),
\end{equation}
leads to the linear potential between heavy quarks, while the essentially 
nonlocal and velocity- and spin-dependent interaction generated by 
nonperturbative QCD exhibits itself as corrections of order of $1/m^2$ to 
the leading linear confinement potential term. These corrections, known as 
Eichten-Feinberg-Gromes relations were derived in several ways, see [1-3].

In the present paper we discuss the approach based on the most 
straightforward use of the Feynman-Schwinger representation for the quark 
Green function [3-6]. The Green function of the quark in the given external 
field is the product of the quadratic Dirac propagator and the linear 
Dirac operator:

\begin{equation}
\frac {1}{m-\hat D}=(m+\hat D)\frac {1}{m^2-\hat {D}^2}
\end{equation}

In the approach under discussion the Feynman-Schwinger representation is 
written out only for the quadratic part of the expression (3), and, as the 
result, the effective Hamiltonian of the $q\bar q$ system contains the 
nonhermitian spin-dependent part. The energy eigenvalues, however, are 
real. Our aim is to study this unusual phenomenon, and to demonstrate 
that the procedure is quite legitimate at least in leading order in 
$1/m^2$, and the resulting interaction is equivalent to the standard one 
[1, 3].

%-----------------------------------------------------------------------------
\section{$q\bar q$ Green function and the effective Lagrangian}
%\begin{center}
%\bf{2. $q\bar q$ Green function and the effective Lagrangian.}
%\end{center}
First we consider the Green function of quark-antiquark system [4].

\begin{equation}
G_{q\bar q}=\left< \Gamma <x|\frac {1}{m_1-\hat D_1}|y>\Gamma <y|\frac {1}{m_2-\hat D_2}|x>\right>_A
\end{equation}
where $\hat D$ is Dirac operator and averaging is performed over a
nonperturbative field A.
In what follows we assume $m_2\to \infty$. For finite $m_2$ the result 
is generalized wihout difficulties.

Each operator $\frac {1}{m-\hat D}$ is represented in the form
$(m+\hat D)\frac {1}{m^2-\hat D^2}$, to exponentiate the positively
defined value of the denominator( Feynman-Schwinger method [5]). For the 
latter we have

\begin{equation}
<x|\frac {1}{m^2-\hat D^2}|y>=\int_0^\infty ds Dz_{\mu} P \exp
\left(-\int_0^s d\tau\left(m^2+\frac {\dot z^2}{4}-g\Sigma F
+igA_{\mu}\dot z_{\mu}\right)\right)
\end{equation}
where $D_{\mu}=\partial_{\mu}-igA_{\mu}$, $P$ is an ordering operator and,  
in the Euclidean space, 

\begin{equation}
\Sigma _{\mu \nu}=\frac {1}{4i}\left[\gamma _{\mu},\gamma _{\nu}\right];~~
\{ \gamma _{\mu },\gamma _{\nu }\}=2\delta _{\mu \nu }.
\end{equation}
Linear operator $\hat D(A)$ corresponds to $-\frac {1}{2}\dot z_{\mu }
(s)\gamma _{\mu }$ in the functional integral, [4, 7].
%++++++++++++++++++++++++++++++++++++++++++++++++++++++++++++++++++++++
Indeed, the matrix element
$< x \mid \frac {-iD_{\mu }}{\hat {D}^2}\mid y >$
can be represented as

$$
   \int\limits^{\infty}_0 d s < x \mid - i D_{\mu} e^{-\hat{D}^2 s}
   \mid y > ~ = $$
$$\int\limits^{\infty}_0 d s < x \mid (p_{\mu}- gA_{\mu})
e^{-(\hat p-g\hat A )^2 s}
   \mid y > ~ , $$
where the linear operator $-iD_{\mu}=
p_{\mu} - g A_{\mu}$ acts on the final state $< x \mid$;
introducing the functional integration over the momenta one has

$$\int\limits^{\infty}_0 d s  \int D z D p (p_{\mu}(s)- gA_{\mu}(s))
e^{-\int\limits^s_0 d \tau \{(\hat p(\tau )-g\hat A(z(\tau)) )^2+ip\dot z\} }~ = $$
$$  =~ \int\limits^{\infty}_0 d s \int D z D p e^{\int\limits^s_0 i p \dot{z}d \tau}
   \frac{1}{2} \frac{\delta}{\delta p_{\mu}} \mid_{\tau=s}
   e^{-\int\limits^s_0 (p-gA)^2 d \tau - \int\limits^s_0 g \Sigma
   F d \tau } . $$
Integration by parts of this expression
yields the replacement
$D_{\mu} \to -\frac{1}{2} \dot{z}_{\mu} \mid_{\tau=s}$~.

%+++++++++++++++++++++++++++++++++++++++++++++++++++++++++++++++++++++++++
So we are spared from the necessity to average 
preexponent factor $\hat D(A)$ at the cost of appearance of $\dot z_{\mu }$ in the
preexponent in the functional integral. Since this linear factor belongs to an
ending point(x or y), we consider the exponent only.

Integration over $p$ gives the kinetic term $m^2+\frac {\dot z^2}{4}$, Wilson
loop $W(C)=P \exp {ig\int dz_{\mu}A_{\mu}}$ and the spin-dependent term, [4]. 

Instead of integrating over the closed contour we
integrate over the area, using the cluster expansion [4]:

\begin{equation}
\left <W(C)\right >_A=\exp \sum_n \frac {1}{2n!}(ig)^n \int d\sigma_{\mu_1\nu_1}...
d\sigma_{\mu_n\nu_n}<<F_{\mu_1 \nu_1}(1)...F_{\mu_n \nu_n}(n)>>
\end{equation}
where $d\sigma_{\mu \nu}=a_{\mu \nu}d\Omega$ is the area element and we run
over all the points with coordinates $w_{\mu}(\beta ,t)$ inside the contour
$w_{\mu}(\beta ,\tau )=z_{\mu }(t)\beta +\bar z_{\mu}(t)(1-\beta )$, $0\le \beta
\le 1$, $d\Omega=rd\beta dt$; $ra_{\mu \nu }=\acute w_{\mu }\dot w_{\nu }-
\acute w_{\nu }\dot w_{\mu }$.

Since $<\Sigma F(x)W(C)>_A=\left.\Sigma _{\mu \nu}\frac {\delta}{\delta
a_{\mu \nu }}\right |_x<W(C)>$, we have exponent $\exp \{\Sigma _{\mu \nu}
\frac {\delta }{\delta a_{\mu \nu }}\}$ which is really a shift operator:

$$a_{\mu \nu }\to \left.a_{\mu \nu }+\Sigma_{\mu \nu }\right |_z$$
 (Operator $\Sigma_{\mu \nu }$ is defined on the trajectory $z$ (or $\bar z)$).
For bilocal correlators we have Wilson loop average expression:

$$<W(C)>=\exp \{(-g^2 \int <<F_{\mu \nu }(w)F_{\alpha \beta }(w^{\prime})>>
a_{\mu \nu }(w)a_{\alpha \beta }(w^{\prime})d\Omega d \Omega^{\prime } )\},$$
and at $r\gg T_g, T\gg T_g$~
$<W(C)>=\exp \{-\sigma \int dtd\beta r\sqrt {a^2/2}\}$,
where $\sigma $ is defined from $\sigma =C \pi g\int\limits_0 ^{\infty }
D(x^2)xdx$ and the constant C and function D are taken from the expression
for bilocal correlator [3]:

$$<<F_{\mu \nu }(w)F_{\alpha \beta }(w^{\prime})>>=
const\left( (\delta _{\mu \alpha }\delta _{\nu \beta }-\delta _{\mu \beta }
\delta _{\nu \alpha })D(h^2)+\right.$$
$$+\left.1/2(\partial _{\alpha }(h_{\mu }\delta _{\beta \nu }
-h_{\nu }\delta _{\beta \mu })
+\partial _{\beta }(h_{\nu }\delta _{\alpha \mu }
-h_{\mu }\delta _{\alpha \nu }))
D_1(h^2)\right ), h=w-w^{\prime}$$.

For spin-orbit term we obtain the expression

\begin{equation}
\frac {\sigma }{2\mu }\frac {\Sigma a}{\sqrt {a^2/2}}=
i\frac {\sigma }{2\mu }\frac {\vec \alpha \vec n+\vec \sigma \vec v_{\perp }}
{\sqrt {1+\vec v_{\perp }^2}}
\end{equation}
in the Euclidean space (Instead of integration over $z_0$ we integrate
over $\mu $ in the functional integral, [6]: $1/(2\mu (z_0))=
\frac {dz_0}{dt}, v_i=\frac {dz_i}{dz_0})$, $\mu $ plays the role of an 
effective mass.

In the Minkovsky space (8) has the form

\begin{equation}
V_{SL}=\frac {\sigma }{2\mu }\frac {i\vec \alpha \vec n+\vec \sigma
\vec v_{\perp}}{\sqrt {1-\vec v_{\perp }^2}}
\end{equation}
and the effective Lagrangian equals to

\begin{equation}
L=-\frac {\mu }{2}(1-\vec v^2)-\frac {m^2}{2\mu }-
\sigma r \int d\beta \sqrt {1-\vec v_{\perp }^2\beta ^2}+
\frac {\sigma }{2\mu }\frac {i\vec \alpha \vec n+\vec \sigma
\vec v_{\perp }}{\sqrt {1-\vec v_{\perp }^2}}
\end{equation}
%-----------------------------------------------------------------------------
In the case of finite mass $m_2$ we get 
$\left[\vec v_{\perp }^{(1)}\beta +\vec v_{\perp }^{(2)}(1-\beta )\right]^2$ 
instead of $\vec v_{\perp }^2$ in the area law, [8], and should replace 
$\vec v_{\perp }$
by $\vec v_{\perp }^{(1,2)}$ for the spin-orbit terms.

%-----------------------------------------------------------------------------
How do these expression change if high order correlators are taken into
account in (7)? At distances $r\gg T_g$ the same expressions are valid and
the only contribution from other correlators is reduced to the renormalization 
of the string tension $\sigma $. Only Kronekker part of correlators( D for
the bilocal one) gives the contribution to $\sigma $ since the
parts with derivatives (like $D_1$) are suppressed at large distances. For small
distances $r\ll T_g$ high order correlators give corrections $(r/T_g)^n$.

The spin-orbit expression also contains $T_g/r$ terms( see
for heavy masses [3])

%\begin{center}
%\bf {3.Effective Hamiltonian.}
%\end{center}
\section{Effective Hamiltonian}

The expression (10) for the Lagrangian makes sense only in the context 
of the Feynman-Schwinger representation, because the last term in (10) contains 
$\gamma $-matrices. To pass to Hamiltonian formulation one has to define 
the canonical momentum as $\vec p=\frac {\partial L}{\partial \vec v}$. 
Clearly, such procedure is possible only for heavy quark, when the last term 
in (10) is treated as perturbation; otherwise the expression for momentum 
would contain $\gamma$-matrices.

%----------------------------------------------------------------------------
For heavy quark the area-law term becomes
$\sigma r \int d\beta \sqrt {1-v_{\perp }^2\beta ^2}\approx
\sigma r (1-\frac {1}{6}v_{\perp }^2)$, integration over $\mu $ gives $\mu =m$, 
the Hamiltonian has a form

\begin{equation}
H=m+\frac {p^2}{2m}+ \sigma r -\frac {\sigma L^2}{6m^2 r} +V_{SL},
\end{equation}

\begin{equation}
V_{SL}=-\frac {\sigma }{2m}\left ( \frac {\vec \sigma \vec L}{mr}+
i\vec \alpha \vec n\right ),
\end{equation}
and we are left with the nonhermitian part in the Hamiltonian. 
The appearance of it is not surprising,
it is caused by the fact that we have exponentiated only the quadratic
part of the Dirac operator, and the projective operator has not been
exponentiated. 

It appeares, however, that the Hamiltonian (11) has real eigenvalues. There 
is nothing mystical in such situation, and the corresponding examples are 
given in the Appendix. In the Hamiltonian (11) $V_{SL}$ is considered as the 
perturbation. Taking the wave function of zero approximation as that of a 
free massive particle ${\phi \choose \frac {\vec \sigma \vec p}{2m}\phi }$, 
we get the matrix element of $i\vec \alpha \vec n $
in the form $\phi ^+\frac {\vec \sigma \vec L}{mr}\phi $.

%*****************************************************************************

It is necessary to work carefully with new wave functions if we want to
apply perturbation theory for new Hamiltonian: new metric norm M is introduced:

$$E^{(1)}\int {\psi ^+}^{(0)}M\psi ^{(0)}=
\int {\psi ^+}^{(0)}HM\psi ^{(0)},$$
 so that $(HM)^{+}=HM$ (the pseudohermitian condition [9])

The Hamiltonians of such kind were considered in detail in the paper [9]. 
According to the theorem in [9] we can transform pseudohermitian Hamiltonian 
with real eigenvalues to the hermitian one, and vice versa, if we obtain a 
hermitian Hamiltonian by a nonunitary transformation we can expect real 
eigenvalues of the original Hamiltonian.

Using Foldy-Wouthuysen transformation, [8]
$$\tilde \psi =\exp {(iS)}\psi , 
\tilde H=\exp {(iS)}H\exp {(-iS)}$$ with $$S=\frac{1}{2m}\vec {\alpha }\vec p,$$ 
we get

\begin{equation}
\tilde H=m+\frac {p^2}{2m}+ \sigma r -\frac {\sigma L^2}{6m^2 r} +\tilde V_{SL}
\end{equation}

\begin{equation}
\tilde V_{SL}=-\frac {\sigma }{2m}\left ( 
\frac {\vec \sigma \vec L}{mr}-\frac {\vec \sigma \vec L}{2mr}+
\frac {1}{2mr}\right )
=\frac {\sigma }{4m^2r}\vec \sigma \vec L+
\frac {\sigma }{4m^2r},
\end{equation}
which is well-known, [1-3,8]. 
The contribution from the electric field reduces the one from the 
magnetic field by the factor 2.
%----------------------------------------------------------------------------
%\begin{center}
%\bf {4. Discussion and conclusions}
%\end{center}
\section{Discussion and conclusions}

The nonunitary Foldy-Wouthuysen transformation which leads to equations (13), 
(14) should be compared with the usual one. If one first applies the usual 
unitary Foldy-Wouthuysen transformation for the quark in the given external 
field, writes out the Feynman-Schwinger representation for the transformed 
Green function and averages over the background field after that, as it was 
done in [3], no problems with nonhermitian Hamiltonian arises, and one 
arrives to the expressions (13), (14) straightforwardly.
%----------------------------------------------------------------------------
It should be noticed that 
the final Hamiltonian and the corresponding equations are not equivalent to 
Dirac equation for one particle in an external field. This is the case since 
after the averaging over nonperturbative field we get nonlocal 
interaction, and the real dynamical object is a string with quarks at the ends. The effective 
field is distributed between quarks and effective string.
The spin-dependent interactions, 
on the contrary, are local and are defined along the contour: it is quark 
(antiquark) which has spin, and feels the dynamics. We have got area law 
and spin-orbit term (14) using successively Feynman-Schwinger formalism  
without introduction of the effective field for a particle.On the other 
hand, if one wishes to consider the problem of a Dirac particle in an 
external field, proceeding from the intermediate result - area law, and, 
in the case of a heavy quark, from the specific form of the string 
correction, which can be treated as external potential 
$\vec A=1/3\vec n\times \vec L$, [10], doing in such a way one should obtain 
with neccessity 1/6 in (14) instead of 1/4, [11]. 
At this point the question arises: why this external potential is treated 
as something that should be substituted into the linear Dirac equation? As 
the original Feynman-Schwinger representation was used only for the 
quadratic part of Dirac operator, this procedure is not well 
grounded at all.\\
We have demonstrated that it is possible to derive the spin-orbit force in 
the Wilson loop context, the results coincide with the well-known ones, and 
the Gromes relations are satisfied. We show, on the other hand, that if the 
particular form of Feynman-Schwinger representation is used, in which the  
projective operator is not included into the path integral, such procedure 
does not allow to go beyond the leading order in $1/m^2$. \\
The auther is grateful for useful discussions to A.Yu. Dubin, 
Yu.S. Kalashnikova, Yu.A. Simonov and acknowledges financial support 
of the Russian Fund for Fundamental Research N. 96-02-19184a.\\[0.5 cm]

%-------------------------------------------------------------------------
\begin{center}
\large{\bf {Appendix. Some examples of pseudohermitian Hamiltonian operator}} 
\end{center}
\setcounter{equation}{0}
\def\theequation{A.\arabic{equation}}

The first example is Klein-Gordon equation for a free heavy-mass particle, [9]:

$$(p_0-p^2-m^2)\phi =0,$$ 
which can be transformed into:
$$(\sqrt {p^2+m^2}+E)(\sqrt {p^2+m^2}-E)\phi =0$$ 
and after expanding the 
root becomes  

$$(m+\frac {p^2}{2m}+E)(m+\frac {p^2}{2m}-E)\phi -\frac {p^4}{4m^2}\phi =0$$

Defining $\chi=\frac {(m+\frac {p^2}{2m}-E)}{-i\frac {p^2}{2m}}\phi $, 
we obtain the equation $ \hat H\psi =E\psi$ for the two-component wave 
function $\psi ={\phi \choose \chi }$, with the following Hamiltonian:

\begin{equation}
\label{A.1}
\hat H=\left (m+\frac {p^2}{2m}\right )\left (\begin{array}{cc}
1& 0\\0& 1\end{array}\right )
+i\frac {p^2}{2m}\left (\begin{array}{cc}
0& 1\\1& 0
\end{array}
\right )
\end{equation}
$\hat H$ has real eigenvalues although it is nonhermitian operator.
For such Hamiltonian another norm M should be defined:

$$\int \psi ^{+}\psi \ne 1,$$ 
$$\int \psi ^{+}M\psi =1.$$
%.....................................................................
The second example is Klein-Gordon equation for a massive particle 
in an external field:

$$\left (P_0^2-\vec P^2-m^2\right )\phi =0,$$

where $P_0=-i\partial _0-A_0$ ,  $P_i=p_i-A_i$.

For new wave function $\tilde \phi =(P_0+\Omega )\phi$ ,
 $\Omega =\sqrt {\vec P^2+m^2}$
we have the equation:

\begin{equation}
\label{A.2}
\left (P_0-\Omega +[\Omega,P_0]\frac {1}{P_0+\Omega }\right )\tilde \phi =0
\end{equation}

For heavy mass 
$\Omega \approx m+\frac {\vec P^2}{2m}-\frac {\vec P^4}{8m^3}$

and 
\begin{equation}
\label{A.3}
\tilde H\approx m+A_0
+\frac {\vec P^2}{2m}-\frac {\vec P^4}{8m^3}
-\frac {1}{4m^2}[\vec P^2,A_0]
+\frac {1}{16m^4}[\vec P^2,[\vec P^2,A_0]]
+\frac {1}{8m^4}[\vec P^4,A_0]
\end{equation}

The first order term is nonhermitian but after the nonunitary 
Foldy-Wouthuysen transformation $\tilde {\tilde {\phi}}
=\exp {iS}\tilde {\phi}$, with $S=-\frac {i}{4m^2}P^2$ 
the new Hamiltonian becomes a hermitian one up to the second order of 1/m:

\begin{equation}
\label{A.4}
\tilde {\tilde H}\approx m+A_0
+\frac {\vec P^2}{2m}-\frac {\vec P^4}{8m^3}
+\frac {1}{32m^4}[\vec P^2,[\vec P^2,A_0]]
\end{equation}

As we can see nonutary transformation was required in order to represent the 
Hamiltonian in a hermitian form.

%...........................................................................

The third example is quadratic Dirac equation for a particle in the 
Coulomb field:

$$(\hat P^2-m^2)\psi =0 .$$

\begin{equation}
\label{A.5}
(P_0^2-\vec p^2-m^2-\Sigma F)\psi =0 .
\end{equation}
Since $P_0\psi =(E-A_0)\psi ,(A_0=\alpha/r , \alpha < \alpha _{crit})$
we have the equation:

$$\left (m+\varepsilon -A_0\right )^2\psi 
=\left (p^2 +m^2+\Sigma F\right )^2\psi $$
So the final equation is $\hat H\psi =E\psi , $
where 

$$\hat H=m+A_0+\frac {p^2}{2m}+\frac {\Sigma F}{2m}-
\frac {(\varepsilon-A_0)^2}{2m}.$$
For a nonrelativistic particle it is possible to replace $\varepsilon -A_0$ 
by $\frac {p^2}{2m}$,and

\begin{equation}
\label{A.6}
\hat H=m+A_0+\frac {p^2}{2m}
-\frac {p^4}{8m^3}+\frac {i\vec \alpha \vec E}{2m}
\end{equation}
The last term is nonhermitian although from the linear Dirac equation we 
have real eigenvalues for energy.\\[0.5cm]

%__---------------------------------------------------------------------------
%\newpage

%----------------------------------------------------------------------------

\end{document}